\documentclass[9pt, twocolumn, technote, twoside]{IEEEtran}

\IEEEoverridecommandlockouts                              	
\usepackage[T1]{fontenc}
\usepackage{graphicx}
\graphicspath{{figures/}}
\usepackage{amssymb,amsmath,amsfonts}
\usepackage{url,changebar,bm,xspace,dsfont}
\usepackage{theorem}
\usepackage{epsfig}
\usepackage{epstopdf}
\usepackage{cite}
\usepackage{verbatim}
\usepackage[usenames,dvipsnames]{color}
\usepackage{flushend}
\usepackage{tikz}

\newtheorem{remark}{Remark}
\newtheorem{definition}{\bfseries Definition}
\newtheorem{theorem}{\bfseries Theorem}
\newtheorem{lemma}{\bfseries Lemma}

\newtheorem{corollary}{\bfseries Corollary}
\newtheorem{example}{\bfseries Example}

\newtheorem{proposition}{\bfseries Proposition}

\makeatother

\def\bnull{\bm{0}}

\def\x{\bm{x}}\def\y{\bm{y}}\def\w{\bm{w}}\def\v{\bm{v}}\def\f{\bm{f}}\def\g{\bm{g}}\def\a{\bm{a}}

\def\bphi{\bm{\varphi}}\def\bpsi{\bm{\psi}}

\begin{document}

\title{Sub-homogeneous positive monotone systems\\ are insensitive to heterogeneous time-varying delays}

\author{Hamid Reza Feyzmahdavian$^{\star}$, Themistoklis Charalambous$^{\star}$, and Mikael Johansson$^{\star}$
\thanks{$^{\star}$Department of Automatic Control, School of Electrical Engineering and ACCESS Linnaeus Center, Royal Institute of Technology (KTH), SE-100 44 Stockholm, Sweden.
Emails: {\tt \{hamidrez, themisc, mikaelj\}@kth.se.}}
}

\maketitle

%
%

\begin{abstract}

We show that a sub-homogeneous positive monotone system with bounded heterogeneous time-varying delays is globally asymptotically stable if and only if the corresponding delay-free system is globally asymptotically stable. The proof is based on an extension of a delay-independent stability result for monotone systems under constant delays by Smith to systems with bounded heterogeneous time-varying delays. Under the additional  assumption of positivity and sub-homogeneous vector fields, we establish the aforementioned delay insensitivity property and derive a novel test for global asymptotic stability. If the system has a unique equilibrium point in the positive orthant, we prove that our stability test is necessary and sufficient. Specialized to positive linear systems, our results extend and sharpen existing results from the literature.

\end{abstract}

%
%

\section{Introduction}\label{sec:intro}

A plethora of real world physical systems involve variables that are positive by nature. Such variables can be the power levels in wireless transmitters, population levels, probabilities and concentrations of substances.  A dynamical system is said to be \textit{positive} if its states are non-negative for all time whenever the initial conditions are non-negative~\cite{Smith:95,Farina:00,Haddad:10}. Due to their wide variety of applications, positive systems have been a subject of much recent attention in engineering and mathematics (see  \textit{e.g.},~\cite{Leenheer:01,2009:Florian,Ruffer:10,Rantzer:11,Lemmens:12,Grussler:12,2012:unconditional,Briat:13,Tanaka:13,Feyzmahdavian:13-0,Rami:14,Ebihara:14,Jonsson:14,
Fornasini:14,Roszak:14,Shen:14,Aleksandrov:14,Li:14,Valcher:14} and references therein).

Models of physical systems are often derived under the assumption that the system evolution depends only on the current values of the state variables. However, in many cases, the system state is also affected by previous values of the states. For example, in distributed systems where exchange of information or materials is involved, delays are inevitable. For this reason, the study of stability and control of dynamical systems with delayed states is essential and of practical importance. In general, time delays limit the performance of closed-loop control systems and may even render an otherwise stable system unstable~\cite{Hale:93}. However, an astonishing property of positive linear systems is that they are insensitive to certain classes of time delays in the following sense: a positive linear system with delays is asymptotically stable if the corresponding delay-free system is asymptotically stable~\cite{Haddad:04,Ngoc:06,Buslowicz:08,
Rami:09,Liu:09,Liu:10,Hamid:13}.

While the asymptotic stability of positive linear systems in the presence of time delays has been thoroughly investigated, the theory for \textit{nonlinear} positive systems is considerably less well-developed (see, \textit{e.g.},~\cite{Mason:09,Vahid:10,Hamid:14TAC, Ngoc:13,Vahid:14,Feyzmahdavian:14-2} for some notable exceptions). The most relevant one is~\cite{Vahid:14} in which it is shown that a class of nonlinear positive systems, called \textit{sub-homogeneous positive monotone systems}, are insensitive to \textit{constant} delays. In practice, however, delays are often \textit{time-varying}. Hence, a natural question is if sub-homogeneous positive monotone systems are insensitive also to time-varying delays. It is reasonable to conjecture that these systems are insensitive to time-varying delays, at least as long as the delays are bounded. However, proving or disproving the conjecture is not trivial. The main reason for this is that the proof technique in~\cite{Vahid:14} relies on a fundamental monotonicity property of trajectories of monotone systems with constant delays, which does not hold when the delays are time-varying.

This paper shows that the conjecture is true. Transforming the stability problem with \textit{heterogeneous} time-varying delays into one with constant delays, we demonstrate that a sub-homogeneous positive monotone system with arbitrary bounded heterogeneous time-varying delays is globally asymptotically stable if and only if the corresponding system without delay is globally asymptotically stable. Since sub-homogeneous positive monotone systems include homogeneous positive monotone systems as a special case, our work also extends the results of \cite{Mason:09,Vahid:10,Hamid:14TAC}. Sub-homogeneous positive monotone systems constitute an important and useful class of nonlinear positive systems, since established models of many physical phenomena fall within this class. For example, most power control algorithms in wireless networks can be analyzed as  sub-homogeneous positive monotone systems~\cite{Yat:95,Hamid:13a,Feyzmahdavian:12,Feyzmahdavian:14}.

The rest of the paper is organized as follows. In Section~\ref{sec:preliminaries}, we introduce the notation and review some preliminaries that are essential for the development of our results.   Section~\ref{sec:Problem Statement} formulates the problem that we address in this paper. Section~\ref{sec:results} presents our main results on the delay-independent stability of monotone systems and the insensitivity of sub-homogeneous positive monotone systems to bounded heterogeneous time-varying delays. Illustrative examples are also included during the development of the results. Finally, conclusions and future directions are given in Section~\ref{sec:conclusions}.

%
%

\section{Notation and Preliminaries}\label{sec:preliminaries}

\subsection{Notation}

Vectors are written in bold lower case letters and matrices in capital letters. We let $\mathbb{R}$ and $\mathbb{N}$ denote the set of real numbers and natural numbers, respectively. The non-negative orthant of the \textit{n}-dimensional real space $\mathbb{R}^n$ is represented by~$\mathbb{R}^n_+$. The $i${th} component of a vector $\x\in \mathbb{R}^n$ is denoted by $x_i$, and the expressions $\x \leq \y$ and $\x < \y$ indicate that $x_i \leq y_i$ and $x_i < y_i$ for all components~$i$, respectively. For a matrix~$A\in \mathbb{R}^{n\times n}$, $a_{ij}$ denotes the entry in row $i$ and column~$j$. A matrix $A\in \mathbb{R}^{n\times n}$ is said to be \textit{non-negative} if $a_{ij}\geq 0$ for all $i$ and~$j$. It is called \textit{Metzler} if $a_{ij}\geq 0$ for all $i\neq j$. For a real interval~$[a,b]$ and an open set $\mathcal{W}\subseteq\mathbb{R}^{n}$, $\mathcal{C}\bigl([a,b],\mathcal{W}\bigr)$ denotes the space of all real-valued continuous functions on $[a,b]$ taking values in $\mathcal{W}$. The upper-right Dini-derivative of a continuous function $h:~\mathbb{R}~\rightarrow~\mathbb{R}$ is denoted by $D^+h(\cdot)$.

\subsection{Preliminaries}

Next, we review the key definitions and results necessary for developing the main results of this paper. We start with the definition of~\textit{cooperative vector fields}.

\begin{definition}
A vector field $\f:\mathcal{W} \rightarrow \mathbb{R}^{n}$ which is continuously differentiable on the open and convex set~$\mathcal{W}\subseteq\mathbb{R}^{n}$ is said to be cooperative if the Jacobian matrix $\frac{\partial \f}{\partial \x}(\a)$ is Metzler for all $\a\in\mathcal{W}$.
\end{definition}
Cooperative vector fields satisfy the following property:

%
%

\begin{proposition}\textup{\textbf{\cite[Remark 3.1.1]{Smith:95}}}
Let $\f:\mathcal{W} \rightarrow \mathbb{R}^{n}$ be cooperative. For any two vectors $\x$ and $\y$ in $\mathcal{W}$ with $x_i=y_i$ and $\x \leq \y$, we have $f_i(\x) \leq f_i(\y)$.
\label{Proposition 1}
\end{proposition}
The following definition introduces~\textit{homogeneous} and~\textit{sub-homogeneous vector fields}.
\begin{definition}
A vector field $\f:\mathbb{R}^{n} \rightarrow \mathbb{R}^{n}$ is called homogeneous of degree $\alpha>0$ if
\begin{align*}
\f(\lambda\x)=\lambda^{\alpha}\f(\x), && \forall \x \in \mathbb{R}^{n},\;\forall \lambda > 0,
\intertext{and it is said to be sub-homogeneous of degree $\alpha>0$ if}
\f(\lambda\x)\leq \lambda^{\alpha}\f(\x), && \forall \x \in \mathbb{R}^{n},\;\forall \lambda\geq 1.
\end{align*}
\end{definition}
It is easy to verify that every homogeneous vector field is also sub-homogeneous. However, the converse is not true; \textit{e.g.},  $f(x)=x+1$ is sub-homogeneous but not homogeneous. Finally, we define \textit{order-preserving vector fields}.
\begin{definition}
A vector field $\g:\mathcal{W} \rightarrow \mathbb{R}^{n}$ is called order-preserving on $\mathcal{W}\subseteq\mathbb{R}^{n}$ if for any $\x,\y\in\mathcal{W}$ such that $\x \leq \y$, it holds that $\g(\x) \leq \g(\y)$.
\end{definition}

%
%

\section{Problem Statement}\label{sec:Problem Statement}

Consider the following nonlinear dynamical system with heterogeneous time-varying delays
\begin{eqnarray}
\hspace{-0.08cm}\left\{ \begin{array}{ll}
\dot{x}_i\bigl(t\bigr)=f_i\bigl(\x(t)\bigr)\\[0.07cm]
\hspace{1.2cm}+g_i\bigl(x_1(t-\tau_1^i(t)),\ldots,x_n(t-\tau_n^i(t))\bigr),&\hspace{-0.22cm} t\geq 0,\\[0.07cm]
x_i\bigl(t\bigr)=\varphi_i\bigl(t\bigr),&\hspace{-1.4cm} t\in[-\tau_{\max},0].
\end{array} \right.
\label{System 1}
\end{eqnarray}
Here, $i\in\{1,\ldots,n\}$, $\x(t)=(x_1(t),\ldots,x_n(t))\in\mathbb{R}^{n}$ is the state vector, $\f(\x)=(f_1(\x),\ldots,f_n(\x))$ and $\g(\x)=(g_1(\x),\ldots,g_n(\x))$ are continuously differentiable vector fields on the open and convex set $\mathcal{W}\subseteq\mathbb{R}^{n}$, and $\bphi(t)=(\varphi_1(t),\ldots,\varphi_n(t))\in \mathcal{C}\bigl([-\tau_{\max},0],\mathcal{W}\bigr)$ is the vector-valued function specifying the initial condition of the system. For all $i$ and $j$, the delays $\tau_j^i(t)$ are continuous with respect to time, and satisfy
\begin{align*}
0\leq \tau_j^i(t)\leq \tau_{\max},\quad \forall t\geq 0.
\end{align*}
Note that the maximum delay bound $\tau_{\max}$ may be unknown, that $\tau_j^i(t)$ are not necessarily continuously differentiable, and that no restriction on their derivative (such as $\dot{\tau}_j^i(t)<1$) is imposed. Since $\bphi(t)$ and $\tau_j^i(t)$ are continuous functions of time, the existence and uniqueness of solutions to~\eqref{System 1} follow from~\cite[Theorem 2.3]{Hale:93}. We denote the unique solution of~\eqref{System 1} corresponding to the initial condition $\bphi(t)$ by $\x(t,\bphi)$. The equilibria of~\eqref{System 1} are constant functions $\bphi(t)=\x^{\star}$, $t\in [-\tau_{\max},0]$, where the vector $\x^{\star}\in\mathcal{W}$ satisfies
\begin{align}
\f(\x^{\star})+\g(\x^{\star})=\bnull.
\label{equilibrium}
\end{align}
In general,~\eqref{equilibrium} may have more than one solution $\x^{\star}$ and, hence, system~\eqref{System 1} may have multiple equilibrium points.

In this paper, we study the delay-independent stability of systems of the form~\eqref{System 1} which are \textit{monotone}:
\begin{definition}
The time-delay system~\eqref{System 1} is called monotone if for any initial conditions $\bphi(t),\bphi'(t)\in \mathcal{C}\bigl([-\tau_{\max},0],\mathcal{W}\bigr)$, $\bphi(t)\leq \bphi'(t)$ for all $t\in[-\tau_{\max},0]$ implies that
\begin{align*}
\x(t,\bphi)\leq \x(t,\bphi'),\quad \forall t\geq 0.
\end{align*}
\end{definition}
Loosely speaking, the trajectories of monotone systems starting at ordered initial conditions preserve the same ordering during the time evolution. Monotonicity of~\eqref{System 1} is readily verified using the next result.

\begin{proposition}\textup{\textbf{\cite[Theorem 5.1.1]{Smith:95}}}\label{Proposition 2}
Suppose that $\f$ is cooperative on $\mathcal{W}$ and $\g$ is order-preserving on $\mathcal{W}$. Then, system~\eqref{System 1} is monotone in~$\mathcal{W}$.
\end{proposition}

System~\eqref{System 1} is said to be positive if for any non-negative initial condition $\bphi(t)\in \mathcal{C}\bigl([-\tau_{\max},0],\mathbb{R}^{n}_+\bigr)$, the corresponding state trajectory will remain in the positive orthant, that is $\x(t,\bphi)\in\mathbb{R}_+^{n}$ for all $t\geq 0$. We provide a necessary and sufficient condition for positivity of monotone systems of the form~\eqref{System 1}.

%
%

\begin{proposition}\label{Proposition 3}
Suppose that $\f$ is cooperative on $\mathbb{R}^n_+$ and $\g$ is order-preserving on $\mathbb{R}^n_+$. Then, the monotone system~\eqref{System 1} is positive if and only if
\begin{align}
\f(\bnull)+\g(\bnull)\geq \bnull.
\label{Proposition 2-1}
\end{align}
\end{proposition}

\begin{IEEEproof}
See Appendix~\ref{Proposition:3}.
\end{IEEEproof}

While the existence of time delays may, in general, induce instability, positive monotone systems whose vector fields are \textit{sub-homogeneous} have been shown to be insensitive to \emph{constant} delays. More precisely, when $\f$ is cooperative and sub-homogeneous on $\mathbb{R}^n_+$ and $\g$ is order-preserving and sub-homogeneous on $\mathbb{R}^n_+$, system~\eqref{System 1} with constant delays $(\tau_j^i(t)=\tau_{\max}$ for all $i$ and $j$ and all $t\geq 0)$ is globally asymptotically stable for all $\tau_{\max}\geq 0$ if and only if the undelayed system $(\tau_{\max}=0)$ is globally asymptotically stable~\cite{Vahid:14}. By global asymptotic stability of a positive system, we mean that its equilibrium in $\mathbb{R}^{n}_+$ is asymptotically stable for all non-negative initial conditions.

It is clear that constant delays is an idealized assumption as time delays are often time-varying in practice. The main objective of this paper is therefore to determine whether sub-homogeneous positive monotone systems are also insensitive to bounded \emph{heterogeneous time-varying} delays.

%
%

\section{Main Results}\label{sec:results}

Having established our notation and problem formulation, we will now present the main contributions of the paper.

\subsection{Monotone Systems}

The following theorem is our first key result, which establishes a sufficient condition for delay-independent stability of monotone systems, not necessarily positive, with bounded heterogeneous time-varying delays.

%
%

\begin{theorem}\label{Theorem 0}

For the time-delay dynamical system~\eqref{System 1}, suppose that $\f$ is cooperative on $\mathcal{W}$ and $\g$ is order-preserving on $\mathcal{W}$. Suppose also that there exist two vectors $\w$ and $\v$ in $\mathcal{W}$ such that $\w\leq \v$ and
\begin{align}
\begin{split}
\f(\w)+\g(\w)&\geq \bnull,\\
\f(\v)+\g(\v)&\leq \bnull.
\end{split}
\label{Theorem 0-1}
\end{align}
Then, if $\x^{\star}\in\mathcal{W}$ is the only equilibrium point of the monotone system~\eqref{System 1} in $[\w,\v]$, then for all bounded heterogeneous time-varying delays,~$\x^{\star}$ is asymptotically stable with respect to initial
conditions satisfying
\begin{align}
\w\leq\bphi(t)\leq \v,\quad \forall t\in[-\tau_{\max},0].
\label{Theorem 0-2}
\end{align}
\end{theorem}

\begin{IEEEproof}
See Appendix~\ref{Theorem:0}.
\end{IEEEproof}

\begin{example}\label{Example:1}
Consider the time-delay system~\eqref{System 1} with
\begin{eqnarray}
\f(x_1,x_2)=\begin{bmatrix}-x_1-1 \\ x_1-x_2(x_2^2-9)+2 \end{bmatrix},\;\g(x_1,x_2)=\begin{bmatrix} 0  \\ x_1
\end{bmatrix}.
\label{Example 1-0}
\end{eqnarray}
Since $\f$ is cooperative on $\mathbb{R}^n$ and $\g$ is order-preserving on~$\mathbb{R}^n$, according to Proposition~\ref{Proposition 2}, system~\eqref{Example 1-0} is monotone on $\mathbb{R}^n$. Note that as
\begin{align*}
\f(0,0)+\g(0,0)=(-1,2)\ngeq(0,0),
\end{align*}
it follows from Proposition~\ref{Proposition 3} that~\eqref{Example 1-0} is not positive. It is easy to see that this system has three equilibrium points:
\begin{align*}
\x^{\star(1)}=(-1,-3),\;\;\x^{\star(2)}=(-1,0),\;\;\x^{\star(3)}=(-1,3).
\end{align*}
Let $\w^{(1)}=(-3,-5)$ and $\v^{(1)}=(1,-1)$. Since
\begin{align*}
\f\bigl(\w^{(1)}\bigr)+\g\bigl(\w^{(1)}\bigr)=\bigl(2,76\bigr)&\geq (0,0),\\
\f\bigl(\v^{(1)}\bigr)+\g\bigl(\v^{(1)}\bigr)=\bigl(-2,-4\bigr)&\leq (0,0),
\end{align*}
and $\x^{\star(1)}$ is the only equilibrium point of~\eqref{Example 1-0} in $[\w^{(1)},\v^{(1)}]$,  it follows from Theorem~\ref{Theorem 0} that for all bounded time-varying delays, $\x^{\star(1)}$ is asymptotically stable with respect to initial conditions satisfying $\w^{(1)}\leq \bphi(t)\leq\v^{(1)}$, $t\in[-\tau_{\max},0]$. Similarly, $\x^{\star(3)}$ is asymptotically stable for initial conditions $\w^{(3)}\leq \bphi(t)\leq\v^{(3)}$, $t\in[-\tau_{\max},0]$, where $\w^{(3)}=(-3,1)$ and $\v^{(3)}=(1,5)$. For example, letting $\tau_1^2(t)= 4 + \sin(t)$, $t\geq 0$, the simulation results shown in Figure~\ref{Fig:1-0} confirm that $\x^{\star(1)}$ and $\x^{\star(3)}$ are indeed locally asymptotically stable.
\begin{figure}[h]
\centering
\includegraphics [width=0.95\columnwidth]{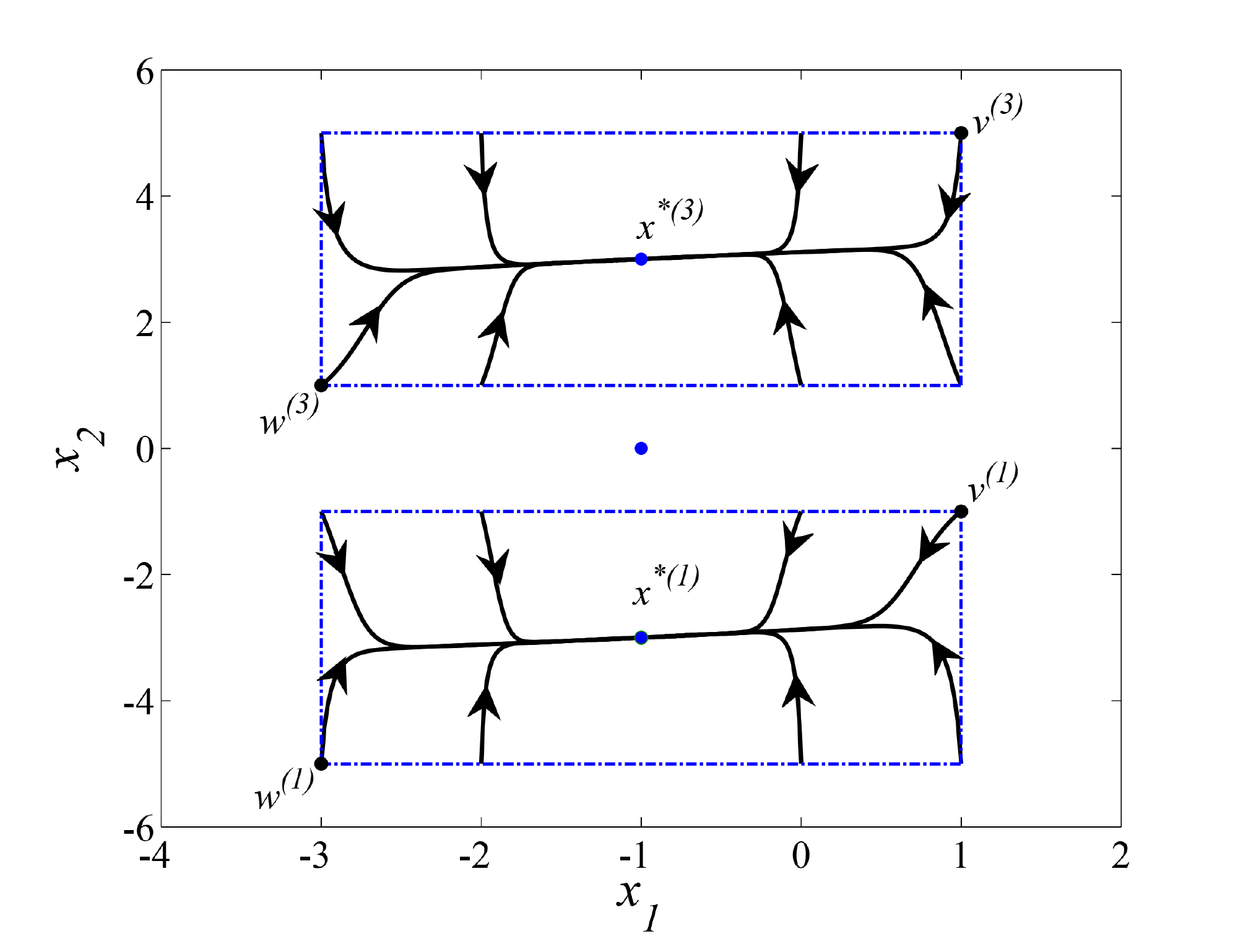}
\vspace{-3.5mm}
\caption{Illustration of asymptotic stability of the monotone system~\eqref{Example 1-0} in Example~\ref{Example:1} under bounded time-varying delays. The equilibrium point $\x^{\star(1)}$ is asymptotically stable with respect to any initial conditions satisfying $(-3,-5)\leq \bphi(t)\leq (1,-1)$,  $t\in[-5,0]$, while $\x^{\star(3)}$ is asymptotically stable for initial conditions $(-3,1)\leq \bphi(t)\leq (1,5)$,  $t\in[-5,0]$.}
\label{Fig:1-0}
\end{figure}
\end{example}

\subsection{Sub-homogeneous Positive Monotone Systems}

Theorem~\ref{Theorem 0} allows us to prove that global asymptotic stability of sub-homogeneous positive monotone systems of the form~\eqref{System 1} is insensitive to bounded heterogeneous time-varying delays.

%
%

\begin{theorem}\label{Theorem 1}

Assume that $\f$ is cooperative on $\mathbb{R}^n_+$ and $\g$ is order-preserving on $\mathbb{R}^n_+$. Furthermore, assume that $\f$ and $\g$ are sub-homogeneous of degree $\alpha>0$. Then, the following statements are equivalent.
\begin{itemize}
\item[(a)] The sub-homogeneous positive monotone system~\eqref{System 1} without delay $(\tau_j^i(t)=0$ for all $i$ and $j$ and all $t\geq 0)$ has a globally asymptotically stable equilibrium point at $\x^{\star}\in \mathbb{R}^n_+$.
\item[(b)] The sub-homogeneous positive monotone system~\eqref{System 1} with arbitrary bounded heterogeneous time-varying delays $\tau_j^i(t)$ has a globally asymptotically stable equilibrium point at $\x^{\star}\in \mathbb{R}^n_+$.
\end{itemize}

\end{theorem}

\begin{IEEEproof}
See Appendix~\ref{Theorem:1}.
\end{IEEEproof}

According to Theorem~\ref{Theorem 1}, the delay-free sub-homogeneous positive monotone system
\begin{align}
\dot{\x}\bigl(t\bigr)=\f\bigl(\x(t)\bigr)+\g\bigl(\x(t)\bigr),\quad t\geq 0,
\label{System 1-0}
\end{align}
is globally asymptotically stable if and only if the equilibrium point at $\x^{\star}\in \mathbb{R}^n_+$ is globally asymptotically stable when arbitrary bounded heterogeneous time-varying delays are introduced into~\eqref{System 1-0}. In other words, global asymptotic stability of~\eqref{System 1-0} implies that of~\eqref{System 1}, and vice versa. This is a significant and surprising property of sub-homogeneous positive monotone systems, since  the existence of time delays may, in general, make a stable system unstable (and, in some special cases, render an unstable system stable).

\begin{remark}
\textup{
The delay-independent stability of monotone systems with \textit{constant} delays was investigated in~\cite{Smith:95}. Using this result,  it has been shown in~\cite{Vahid:10} that \textit{homogeneous} positive monotone systems are insensitive to \textit{constant} time delays. It is clear that\textit{ bounded heterogeneous time-varying} delays include constant delays as a special case. Moreover, every homogeneous vector field is also \textit{sub-homogeneous}. Hence, Theorem~\ref{Theorem 0} extends the result in~\cite{Smith:95} to bounded heterogeneous time-varying delays and Theorem~\ref{Theorem 1} recovers the delay independence of homogeneous positive monotone systems as a special case.
}
\end{remark}

The next lemma, which is instrumental for the proof of Theorem~\ref{Theorem 1},  establishes a necessary condition for the global asymptotic stability of general positive monotone systems (not necessarily sub-homogeneous) with bounded heterogeneous time-varying delays.

%
%

\begin{lemma}\label{Lemma 2}
For the time-delay dynamical system~\eqref{System 1}, suppose that $\f$ is cooperative on $\mathbb{R}^{n}_+$ and $\g$ is order-preserving on~$\mathbb{R}^{n}_+$. If the positive monotone system~\eqref{System 1} has a globally asymptotically stable equilibrium at $\x^{\star}\in \mathbb{R}^n_+$, then the following statements hold:
\begin{itemize}
\item[(a)] There does not exist a vector $\w\neq \x^{\star}$ such that $\w\geq \x^{\star}$ and
\begin{align}
\f(\w)+\g(\w)\geq \bnull.\label{Lemma 2-1}
\end{align}
\item[(b)] There exists a vector $\v>\bnull$ such that $\v>\x^{\star}$ and
\begin{align}
\f(\v)+\g(\v)< \bnull.\label{Lemma 2-2}
\end{align}
\end{itemize}
\end{lemma}

\begin{IEEEproof}
See Appendix~\ref{Lemma:2}.
\end{IEEEproof}

Lemma~\ref{Lemma 2} provides a test for the global asymptotic stability of positive monotone systems of the form~\eqref{System 1} with bounded heterogeneous time-varying delays: if we can demonstrate the existence of a vector $\w\geq \x^{\star}$ satisfying~\eqref{Lemma 2-1} or prove there is no positive vector $\v>\x^{\star}$ satisfying~\eqref{Lemma 2-2}, then the equilibrium at $\x^{\star}$ cannot be globally asymptotically stable.

\begin{example}
Consider the time-delay dynamical system described by~\eqref{System 1} with
\begin{align}
\f(x_1,x_2)=\begin{bmatrix}-x_1^2+x_2 \\ -x_2 \end{bmatrix},\quad  \g(x_1,x_2)=\begin{bmatrix} x_2 \\ x_1
\end{bmatrix}.
\label{Example 1}
\end{align}
\begin{figure}[h]
\centering
\includegraphics [width=0.75\columnwidth]{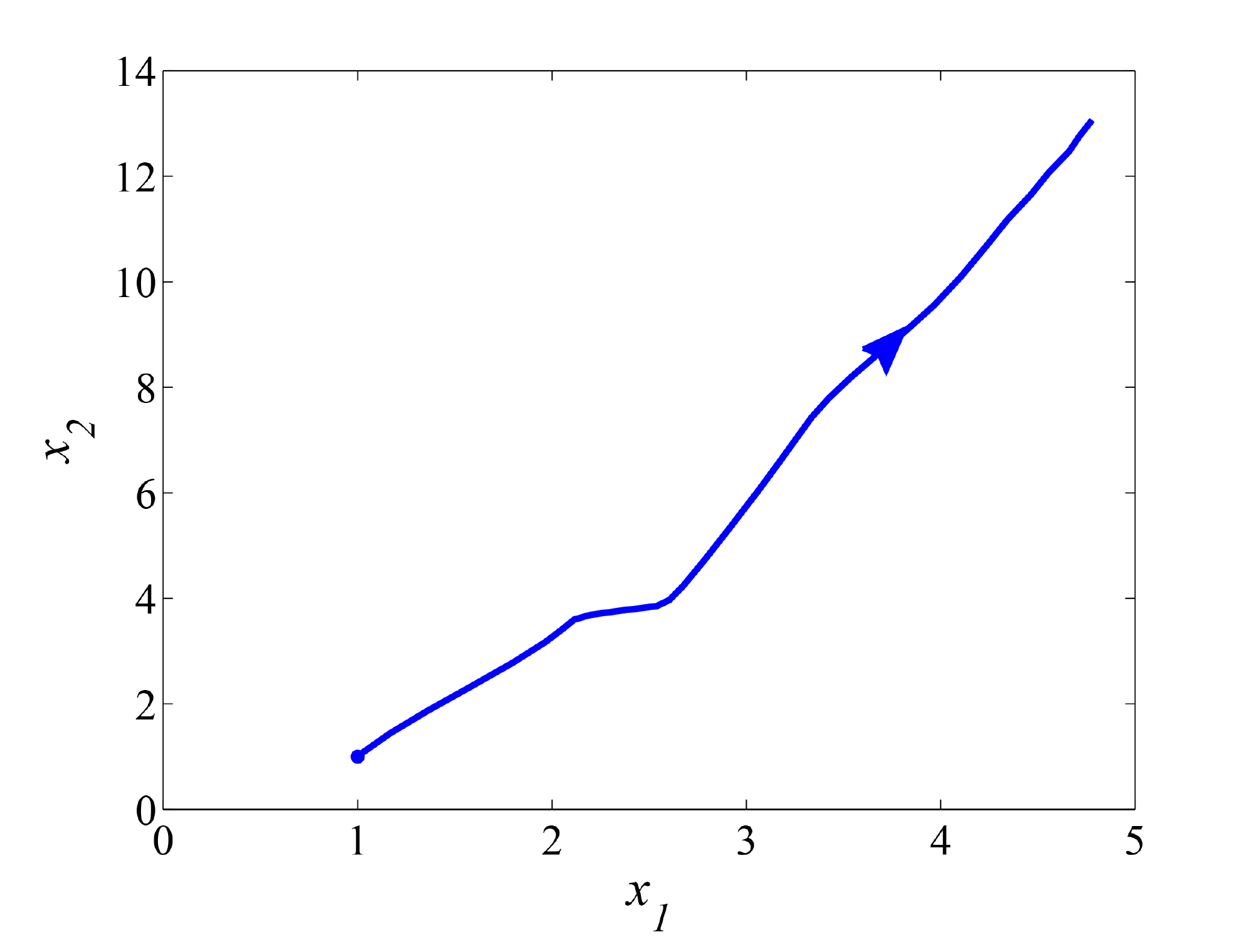}
\vspace{-3.5mm}
\caption{Illustration of a trajectory of the positive monotone system~\eqref{Example 1} corresponding to the initial condition $\bphi(t)=(1,1)$, $\forall t\in[-3,0]$\vspace{-2mm}.} \label{Fig:1}
\end{figure}
It is easy to verify that $\f$ is cooperative on $\mathbb{R}^{n}_+$, $\g$ is order-preserving on~$\mathbb{R}^{n}_+$, and $\f(0,0)+\g(0,0)=(0,0)$. Thus,~\eqref{Example 1} is a positive monotone system with an equilibrium point at the origin. Since
\begin{align*}
\f(1,1)+\g(1,1)=(1,0)\geq (0,0),
\end{align*}
it follows from Lemma~\ref{Lemma 2} that for any bounded heterogeneous time-varying delays, the origin is not a globally asymptotically stable equilibrium of~\eqref{Example 1}. For example, we take $\tau_j^i(t)=2+\sin(t)$, $i,j=1,2$, $ t\geq 0$, and the simulation result is shown in Figure~\ref{Fig:1}, from which one can see that the trajectory of~\eqref{Example 1} starting from the initial condition $\bphi(t)=(1,1)$, $\forall t\in[-3,0]$, does not converge to the origin.

\end{example}

\begin{remark}
\textup{Previous works in the literature established necessary conditions for the global asymptotic stability of positive monotone systems without time delays~\cite{Ruffer:10,Vahid:11}. Lemma~\ref{Lemma 2}, therefore, is an extension of these results to delayed positive monotone systems of the form~\eqref{System 1}.}
\end{remark}

The following example illustrates that the necessary conditions given in Lemma~\ref{Lemma 2} are, in general, not sufficient.

\begin{example}
Consider the time-delay system~\eqref{System 1} with
\begin{align}
\f(x_1,x_2)=\begin{bmatrix}-\frac{x_1}{1+x_1^3} \\ -x_2^4 \end{bmatrix},\quad  \g(x_1,x_2)=\begin{bmatrix}  x_2 \\ 0
\end{bmatrix}.
\label{Example 2}
\end{align}
Let the time-delay be given by $\tau_2^1(t)=5-\cos(t)$, $t\geq 0$. It can be easily checked that~\eqref{Example 2} is a positive monotone system with an equilibrium at the origin. Since no non-zero vector $\w\geq \bnull$ satisfying~\eqref{Lemma 2-1} exists~\cite[Example 3.11]{Ruffer:10} and
\begin{align*}
\f(1,\frac{1}{4})+\g(1,\frac{1}{4})=(-\frac{1}{4},-\frac{1}{256})< (0,0),
\end{align*}
\begin{figure}[h]
\centering
\includegraphics [width=0.75\columnwidth]{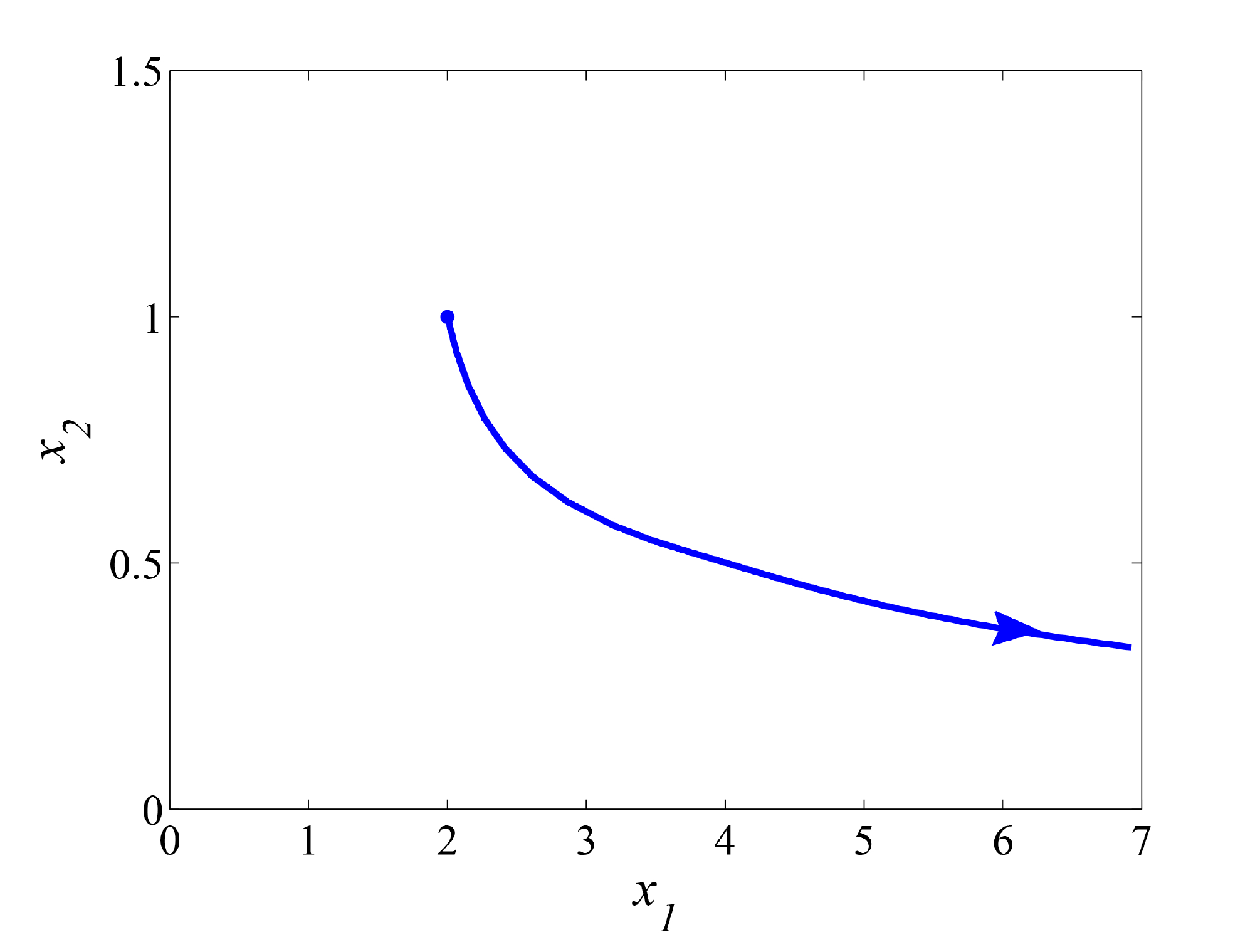}
\vspace{-3.5mm}
\caption{Illustration of a trajectory of the positive monotone system~\eqref{Example 2} corresponding to the initial condition $\bphi(t)=(2,1)$, $\forall t\in[-6,0]$\vspace{-2mm}.} \label{Fig:2}
\end{figure}
the necessary conditions stated in Lemma~\ref{Lemma 2} holds. However, Figure~\ref{Fig:2} shows that the trajectory of~\eqref{Example 2} corresponding to the initial condition $\bphi(t)=(2,1)$, $\forall t\in[-6,0]$, does not converge to the origin, which means that the positive monotone system~\eqref{Example 2} is not globally asymptotically stable. Thus, the necessary conditions in  Lemma~\ref{Lemma 2} are not sufficient.

\end{example}

Next, we show that when a sub-homogeneous positive monotone system of the form~\eqref{System 1} has a unique equilibrium point in~$\mathbb{R}^n_+$, the necessary conditions in Lemma~\ref{Lemma 2} are also sufficient.

%
%

\begin{corollary}\label{Theorem 2}

Assume that $\f$ is cooperative on $\mathbb{R}^n_+$ and $\g$ is order-preserving on $\mathbb{R}^n_+$. Furthermore, assume that $\f$ and $\g$ are sub-homogeneous of degree $\alpha>0$. If $\x^{\star}$ is the only equilibrium of the positive system~\eqref{System 1} in~$\mathbb{R}^n_+$, then the sub-homogeneous positive monotone system~\eqref{System 1} is globally asymptotically stable for any arbitrary bounded heterogeneous time-varying delays if and only if there exists a vector $\v>\bnull$ such that $\v>\x^{\star}$ and
\begin{align*}
\f(\v)+\g(\v)< \bnull.
\end{align*}

\end{corollary}

\begin{example}

Consider the time-delay system~\eqref{System 1} with
\begin{align}
\f(x_1,x_2)=\begin{bmatrix}-2x_1+\frac{x_2}{x_2+2} \\ -2x_2+\frac{x_1}{x_1+2} \end{bmatrix},\quad  \g(x_1,x_2)=\begin{bmatrix} x_1  \\ x_2
\end{bmatrix}.
\label{Example 4}
\end{align}
It is straightforward to verify that~\eqref{Example 4} is a sub-homogeneous positive monotone system. Moreover, this system has two equilibrium points, one is $\x^{\star(1)}=(0,0)$, and the other is $\x^{\star(2)}=(-1,-1)$. Since the origin is the unique equilibrium in~$\mathbb{R}^n_+$, it follows from Corollary~\ref{Theorem 2} that for all non-negative initial conditions and for any bounded heterogeneous time-varying delays, $\x^{\star(1)}$ is globally asymptotically stable.

\end{example}

\subsection{A Special Case: Positive Linear Systems}

We now discuss the delay-independent stability of a special case of~\eqref{System 1}, namely positive linear system of the form
\begin{eqnarray}
 \left\{
\begin{array}[l]{ll}
\dot{\x}\bigl(t\bigr)=A\x\bigl(t\bigr)+B\x\bigl(t-\tau(t)\bigr),\quad t\geq 0,\\[1mm]
\x\bigl(t\bigr)=\bphi\bigl(t\bigr), \quad  \hspace{1.3cm} t\in[-\tau_{\max},0],
\end{array}
\right.
\label{System 2}
\end{eqnarray}
where $A\in\mathbb{R}^{n\times n}$ is Metzler and $B\in\mathbb{R}^{n\times n}$ is non-negative. In terms of \eqref{System 1}, $\f(\x)=A\x$, $\g(\x)=B\x$, and $\tau_j^i(t)=\tau(t)$ for all $i$ and $j$. It is easy to verify that the positive linear system~\eqref{System 2} is a sub-homogeneous positive monotone system with an equilibrium at the origin. Since $A+B$ is Metzler, it follows from~\cite[Proposition 2]{Rantzer:11} that the undelayed system
\begin{eqnarray*}
\dot{\x}\bigl(t\bigr)=(A+B)\x\bigl(t\bigr),
\end{eqnarray*}
is globally asymptotically stable if and only if the following set of linear inequalities in $\v$
\begin{align}
\begin{cases}
\bigl(A+B\bigr)\v&<\bnull,\\
\hspace{1cm}\v&>\bnull,
\end{cases}
\label{LP}
\end{align}
is feasible. This shows that the existence of vector $\v$ satisfying~\eqref{LP} is a necessary and sufficient condition for the asymptotic stability of the positive linear system~\eqref{System 2} with no delays. We have the following special case of Theorem~\ref{Theorem 1}.

\begin{corollary}\label{Theorem 3}

Consider the positive linear system~\eqref{System 2} where $A$ is Metzler and $B$ is non-negative. Then, the following statements are equivalent.
\begin{itemize}
\item[(a)] The linear programming problem~\eqref{LP} has a feasible solution $\v$.
\item[(b)] The positive linear system~\eqref{System 2} without delay is globally asymptotically stable.
\item[(c)] The positive linear system~\eqref{System 2} with any arbitrary bounded time-varying delay is globally asymptotically stable.
\end{itemize}
\end{corollary}

\begin{remark}
\textup{
In~\cite{Liu:10}, it was shown that the positive linear system~\eqref{System 2} is asymptotically stable for \textit{all} bounded time-varying delays if and only if there exists a vector $\v$ satisfying~\eqref{LP}. This result does not allow to conclude stability of the undelayed positive system from the stability of the corresponding delayed system under \textit{some} delays (not under all bounded delays). In contrast, Corollary~\ref{Theorem 3} shows that asymptotic stability of~\eqref{System 2} under \emph{any arbitrary} bounded time-varying delay implies the asymptotic stability of the corresponding undelayed system. Therefore, Corollary~\ref{Theorem 3} is stronger than the result in~\cite{Liu:10}.
}
\end{remark}

%
%

\section{Conclusions and Future Directions}\label{sec:conclusions}

We extended delay-independent stability results for sub-homogeneous positive monotone systems to allow for heterogeneous time-varying delays. Specifically, we proved that a sub-homogeneous positive monotone system is globally asymptotically stable for any bounded heterogeneous time-varying delay if and only if the corresponding delay-free system is globally asymptotically stable. Homogeneous positive monotone systems and positive linear systems constitute special cases. Illustrative examples demonstrate the validity of our results. Extensions to more general classes of positive monotone systems, for which the sub-homogeneity assumption does not hold, is part of ongoing research.

%
%

\appendix

%
%

\subsection{A Technical Lemma}\label{Lemma}

The following lemma plays a key role in this paper.

\begin{lemma}\label{Lemma 3}

Consider the following time-delay dynamical system with constant delays, closely related to system~\eqref{System 1}\textup{:}
\begin{align}
\left\{
\begin{array}[l]{ll}
\dot{\y}\bigl(t\bigr)=\f\bigl(\y(t)\bigr)+\g\bigl(\y(t-\tau_{\max})\bigr),\quad \hspace{0.5cm} t\geq 0,\\[0.05cm]
\y\bigl(t\bigr)=\bpsi\bigl(t\bigr),\quad \hspace{2.2cm} t\in[-\tau_{\max},0].
\end{array}
\right.
\label{System 1-1}
\end{align}
Here, $\f$ is cooperative on $\mathcal{W}$, $\g$ is order-preserving on $\mathcal{W}$, and $\tau_{\max}$ equals the upper bound of the delays $\tau_j^i(t)$ in~\eqref{System 1}.
\begin{enumerate}
\item Assume that there exists a vector $\v\in\mathcal{W}$ satisfying
\begin{align}
\f(\v)+\g(\v)&\leq \bnull,
\label{Lemma 3-1}
\end{align}
and that the initial conditions for systems~\eqref{System 1} and~\eqref{System 1-1} are
$\bphi_v(t)=\v$ and $\bpsi_v(t)=\v$, $t\in[-\tau_{\max},0]$, respectively. Then, the solution $\x(t,\bphi_v)$ to~\eqref{System 1} starting from~$\bphi_v(t)$ satisfies
\begin{align*}
\x(t,\bphi_v)\leq\y(t,\bpsi_v), \quad \forall t\geq 0,
\end{align*}
where $\y(t,\bpsi_v)$ is the solution to~\eqref{System 1-1} with the initial condition~$\bpsi_v(t)$.

\item Assume that there exists a vector $\w\in\mathcal{W}$ satisfying
\begin{align}
\f(\w)+\g(\w)&\geq \bnull,
\label{Lemma 3-2}
\end{align}
and that the initial conditions for systems~\eqref{System 1} and~\eqref{System 1-1} are
$\bphi_w(t)=\w$ and $\bpsi_w(t)=\w$, $t\in[-\tau_{\max},0]$, respectively. Then, it holds that
\begin{align*}
\y(t,\bpsi_w)\leq \x(t,\bphi_w), \quad \forall t\geq 0,
\end{align*}
where $\x(t,\bphi_w)$ and $\y(t,\bpsi_w)$ are solutions to~\eqref{System 1} and~\eqref{System 1-1}, respectively.

\end{enumerate}

\end{lemma}

\begin{IEEEproof}

\textbf{Part 1)} Let $\v\in\mathcal{W}$ be a vector satisfying~\eqref{Lemma 3-1}, and let $\y(t,\bpsi_v)$ be the solution to~\eqref{System 1-1} with respect to the initial condition $\bpsi_v(t)=\v$, $t\in[-\tau_{\max},0]$. Consider the following system
with heterogeneous time-varying delays
\begin{align}
\hspace{-0.28cm}\left\{ \begin{array}{ll}
\dot{x}_i\bigl(t\bigr)=f_i\bigl(\x(t)\bigr)\\[0.07cm]
\hspace{0.8cm}+g_i\bigl(x_1(t-\tau_1^i(t)),\ldots,x_n(t-\tau_n^i(t))\bigr)-\frac{1}{k},&\hspace{-0.2cm} t\geq 0,\\[0.07cm]
x_i\bigl(t\bigr)=\varphi_i\bigl(t\bigr),&\hspace{-1.4cm} t\in[-\tau_{\max},0],
\end{array} \right.
\label{Lemma 3-3}
\end{align}
where $k\in\mathbb{N}$. Let $\x^{(k)}(t,\bphi_v)$ be the solution to~\eqref{Lemma 3-3} with the initial condition $\bphi_v(t)=\v$, $t\in[-\tau_{\max},0]$. Clearly,
\begin{align*}
\x^{(k)}(0,\bphi_v)=\v\leq \y(0,\bpsi_v)=\v.
\end{align*}
We claim that $\x^{(k)}(t,\bphi_v)\leq \y(t,\bpsi_v)$ for all $t\geq 0$. If the result were false, we can assume that there exist an index $m\in\{1,\ldots,n\}$ and a time $t_1\geq 0$ such that
\begin{align}
\begin{split}
\x^{(k)}(t,\bphi_v)&\leq \y(t,\bpsi_v),\quad \forall t\in[0,t_1],\\
x^{(k)}_m(t_1,\bphi_v)&=y_m(t_1,\bpsi_v),
\end{split}
\label{Lemma 3-4}
\end{align}
and
\begin{align}
&\hspace{-1.5cm}D^+x^{(k)}_m(t_1,\bphi_v)\geq D^+y_m(t_1,\bpsi_v).\label{Lemma 3-5}
\end{align}
Since $\f$ is cooperative,  Proposition~\ref{Proposition 1} and~\eqref{Lemma 3-4} imply that
\begin{align}
f_m\bigl(\x^{(k)}(t_1,\bphi_v)\bigr)&\leq f_m\bigl(\y(t_1,\bpsi_v)\bigr).\label{Lemma 3-6}
\end{align}
As $t_1-\tau_j^m(t_1)\in [-\tau_{\max},t_1]$ for all $j\in\{1,\ldots,n\}$ and $\x^{(k)}(t,\bphi_v)= \y(t,\bpsi_v)=\v$ for all $t\in~[-\tau_{\max},0]$, it follows from~\eqref{Lemma 3-4} that
\begin{align}
x^{(k)}_j\bigl(t_1-\tau_j^m(t_1),\bphi_v\bigr)\leq y_j\bigl(t_1-\tau_j^m(t_1),\bpsi_v\bigr),
\label{Lemma 3-7}
\end{align}
irrespectively of whether $t_1-\tau_j^m(t_1)$ is non-negative or not. On the other hand,  it follows from~\cite[Corollary 5.2.2]{Smith:95} and~\eqref{Lemma 3-1} that $\y(t,\bpsi_v)$ is non-increasing for all $t\geq 0$. Thus, for each $j$ we have
\begin{align}
y_j\bigl(t_1-\tau_j^m(t_1),\bpsi_v\bigr)&\leq y_j\bigl(t_1-\tau_{\max},\bpsi_v\bigr),
\label{Lemma 3-8}
\end{align}
where we have used the fact that $\tau_j^m(t_1)\leq \tau_{\max}$ to get the inequality. Since $\g$ is order-preserving, it follows from~\eqref{Lemma 3-7} and~\eqref{Lemma 3-8} that
\begin{align}
g_m\bigl(x^{(k)}_1(t_1-&\tau_1^m(t_1),\bphi_v),\ldots,x^{(k)}_n(t_1-\tau_n^m(t_1),\bphi_v)\bigr)\nonumber\\
&\leq g_m\bigl(y_1(t_1-\tau_{\max},\bpsi_v),\ldots,y_n(t_1-\tau_{\max},\bpsi_v)\bigr)\nonumber\\
&=g_m\bigl(\y(t_1-\tau_{\max},\bpsi_v)\bigr).
\label{Lemma 3-9}
\end{align}
By~\eqref{Lemma 3-6} and~\eqref{Lemma 3-9}, the upper-right Dini-derivative of $x^{(k)}_m(t,\bphi_v)$ along the trajectories of~\eqref{Lemma 3-3} at $t=t_1$ satisfies
\begin{align*}
D^+&x^{(k)}_m(t_1,\bphi_v)\\
&=f_m\bigl(\x^{(k)}(t_1,\bphi_v)\bigr)\\
&\hspace{-0.5cm}+g_m\bigl(x_1(t_1-\tau_1^m(t_1),\bphi_v),\ldots,x_n(t_1-\tau_n^m(t_1),\bphi_v)\bigr)-\frac{1}{k}\\
&\leq f_m\bigl(\y(t_1,\bpsi_v)\bigr)+g_m\bigl(\y(t_1-\tau_{\max},\bpsi_v)\bigr)-\frac{1}{k}\\
&=  D^+y_m(t_1,\bpsi_v)-\frac{1}{k}\\
&< D^+y_m(t_1,\bpsi_v),
\end{align*}
which contradicts~\eqref{Lemma 3-5}. Therefore,
\begin{align}
\x^{(k)}(t,\bphi_v)\leq \y(t,\bpsi_v),\quad \forall t\geq 0.
\label{Lemma 3-10}
\end{align}
Since $k$ was an arbitrary natural number,~\eqref{Lemma 3-10} holds for all $k\in\mathbb{N}$. By letting $k\rightarrow \infty$, $\x^{(k)}(t,\bphi_v)$ converges to the solution $\x(t,\bphi_v)$ of~\eqref{System 1} uniformly on $[-\tau_{\max},\infty)$~\cite[Theorem 2.2]{Hale:93}. This shows that $\x(t,\bphi_v)\leq \y(t,\bpsi_v)$ for all $t\geq 0$.

\textbf{Part 2)} Now, let $\w\in\mathcal{W}$ be a vector satisfying~\eqref{Lemma 3-2}, and let $\y(t,\bpsi_w)$ be the solution to~\eqref{System 1-1} starting from the initial condition $\bpsi_w(t)=\w$, $t\in[-\tau_{\max},0]$. According to~\cite[Corollary 5.2.2]{Smith:95}, $\y(t,\bpsi_w)$ is non-decreasing for all $t\geq 0$. The rest of the proof is similar to the one for Part 1) and thus omitted.

\end{IEEEproof}

%
%

\subsection{Proof of Proposition~\ref{Proposition 3}}\label{Proposition:3}

Assume that $\f(\bnull)+\g(\bnull)\geq \bnull$ and let $\bphi_0(t)$ be the initial condition satisfying $\bphi_0(t)=\bnull$, $t\in[-\tau_{\max},0]$. Since $\f$ is cooperative on $\mathbb{R}^n_+$ and $\g$ is order-preserving on $\mathbb{R}^n_+$, it follows from Proposition~\ref{Proposition 2} that system~\eqref{System 1} is monotone. Thus, for any initial condition $\bphi(t)$ satisfying $\bphi_0(t)\leq \bphi(t)$, $\forall t\in[-\tau_{\max},0]$, it holds that
\begin{align}
\x(t,\bphi_0)\leq \x(t,\bphi),\quad \forall t\geq 0.
\label{Proposition 2-2}
\end{align}
Let $\y(t,\bpsi_0)$ be the solution to system~\eqref{System 1-1} starting from the initial condition $\bpsi_0(t)=\bnull$, $t\in[-\tau_{\max},0]$. It follows from~\cite[Corollary 5.2.2]{Smith:95} and~\eqref{Proposition 2-1} that $\y(t,\bpsi_0)$ is non-decreasing, \textit{i.e},
\begin{align}
\bnull=\bpsi_0(0)\leq\y(t,\bpsi_0),\quad \forall t\geq 0.
\label{Proposition 2-3}
\end{align}
Since, according to Lemma~\ref{Lemma 3}, $\y(t,\bpsi_0)\leq \x(t,\bphi_0)$ for all $t\geq 0$, it follows from~\eqref{Proposition 2-2} and~\eqref{Proposition 2-3} that $\bnull\leq \x(t,\bphi)$ for all $t\geq 0$. Therefore, system~\eqref{System 1} is positive.

Conversely, assume that~\eqref{System 1} is positive. Suppose, for contradiction, that there is an index $m\in\{1,\ldots,n\}$ such that $f_m(\bnull)+g_m(\bnull)<0$. Then,
\begin{align*}
D^+x_m(0,\bphi_0)=f_m(\bnull)+g_m(\bnull)<0,
\end{align*}
and hence there is some $\delta > 0$ such that
\begin{align*}
x_m(t,\bphi_0)< x_m(0,\bphi_0) =0,\quad  \forall t\in (0,\delta).
\end{align*}
Thus, $\x(t)\notin \mathbb{R}^n_+$ for $t\in (0,\delta)$, which is a contradiction.

%
%

\subsection{Proof of Theorem~\ref{Theorem 0}}\label{Theorem:0}

Let $\w$ and $\v$ be vectors such that $\w\leq\v$ and that~\eqref{Theorem 0-1} holds. Define $\bphi_w(t)=\w$ and $\bphi_v(t)=\v$, $t\in[-\tau_{\max},0]$. Since $\f$ is cooperative and $\g$ is order-preserving, according to Proposition~\ref{Proposition 2}, system~\eqref{System 1} is monotone. Thus, for any initial condition $\bphi(t)$ satisfying~\eqref{Theorem 0-2}, we have
\begin{align*}
\x(t,\bphi_w)\leq\x(t,\bphi)\leq \x(t,\bphi_v),\quad \forall t\geq 0.
\end{align*}
Define $\bpsi_w(t)=\w$ and $\bpsi_v(t)=\v$, $t\in[-\tau_{\max},0]$. Let $\y(t,\bpsi_w)$ and $\y(t,\bpsi_v)$ be solutions to system~\eqref{System 1-1} starting from $\bpsi_w(t)$ and $\bpsi_v(t)$, respectively. According to Lemma~\ref{Lemma 3}, $\y(t,\bpsi_w)\leq\x(t,\bphi_w)$ and $\x(t,\bphi_v)\leq\y(t,\bpsi_v)$ for all $t\geq 0$, implying that
\begin{align}
\y(t,\bpsi_w)\leq\x(t,\bphi)\leq\y(t,\bpsi_v),\quad \forall t\geq 0.
\label{Theorem 0-3}
\end{align}
Moreover, according to~\cite[Corollary 5.2.2]{Smith:95}, $\y(t,\bpsi_w)$ is non-decreasing and $\y(t,\bpsi_v)$ is non-increasing for $t\geq 0$, which together with the monotonicity of~\eqref{System 1-1} imply that
\begin{align*}
\w\leq\y(t,\bpsi_w)\leq\y(t,\bpsi_v)\leq \v,\quad \forall t\geq 0.
\end{align*}
Thus, both $\y(t,\bpsi_w)$ and $\y(t,\bpsi_v)$ are bounded and monotone. It now follows from~\cite[Theorem 1.2.1]{Smith:95} that $\y(t,\bpsi_w)$ and $\y(t,\bpsi_v)$ converge to an equilibrium of~\eqref{System 1-1} in $[\w,\v]$, which must be $\x^{\star}$, \textit{i.e}.,
\begin{align}
\lim_{t\rightarrow \infty}\y(t,\bpsi_w)=\lim_{t\rightarrow \infty}\y(t,\bpsi_v)=\x^{\star}.
\label{Theorem 0-4}
\end{align}
It follows from~\eqref{Theorem 0-3} and~\eqref{Theorem 0-4} that $\lim_{t\rightarrow \infty}\x(t,\bphi)=\x^{\star}$. This completes the proof.

%
%

\subsection{Proof of Theorem~\ref{Theorem 1}}\label{Theorem:1}

We first prove that~$(a)$ implies~$(b)$.

$(a)\Rightarrow (b):$ Assume that the sub-homogeneous positive monotone system~\eqref{System 1} without delay, given by
\begin{align}
\dot{\x}(t)=\f(\x(t))+\g(\x(t)),\label{Theorem 1-0-0}
\end{align}
has a globally asymptotically stable equilibrium at $\x^{\star}\in\mathbb{R}^n_+$. Clearly, $\x^{\star}$ is the only equilibrium in $\mathbb{R}^n_+$. Since~\eqref{Theorem 1-0-0} is positive, according to Proposition~\ref{Proposition 3}, we have
\begin{align}
\f(\bnull)+\g(\bnull)\geq\bnull.
\label{Theorem 1-0-1}
\end{align}
Moreover, as $\f+\g$ is cooperative on $\mathbb{R}^n_+$, it follows from~\cite[Proposition 4.2]{Vahid:11} that there is $\v>\bnull$ with~$\v>\x^{\star}$ such that $\f(\v)+\g(\v)< \bnull$. This together with sub-homogeneity of $\f$ and $\g$ implies that for any real constant $\gamma$ with $\gamma\geq 1$,
\begin{align}
\f({\gamma\v})+\g({\gamma\v})&\leq \gamma^\alpha\bigl(\f(\v)+\g(\v)\bigr)< \bnull.
\label{Theorem 1-0-2}
\end{align}
It now follows from Theorem~\ref{Theorem 0},~\eqref{Theorem 1-0-1} and~\eqref{Theorem 1-0-2} that~$\x^{\star}$ is asymptotically stable for any initial conditions satisfying
\begin{align}
\bnull\leq\bphi(t)\leq {\gamma\v},\quad t\in[-\tau_{\max},0].
\label{Theorem 1-0-3}
\end{align}
Let $\bphi(t)\in \mathcal{C}\bigl([-\tau_{\max},0],\mathbb{R}^{n}_+\bigr)$ be an arbitrary non-negative initial condition. As $\v>\bnull$ and $\bphi(t)$ is continuous and, hence, bounded on $[-\tau_{\max},0]$, there exists ${\gamma}\geq 1$ sufficiently large such that~\eqref{Theorem 1-0-3} holds. Therefore, system~\eqref{System 1} with arbitrary bounded heterogeneous time-varying delays is globally asymptotically stable for all non-negative initial conditions.

We next show that~$(b)$ implies~$(a)$.

$(b)\Rightarrow (a):$ Assume that system~\eqref{System 1} with arbitrary bounded heterogeneous time-varying delays $\tau_j^i(t)$ has a globally asymptotically stable equilibrium at $\x^{\star}\in\mathbb{R}^n_+$. Since~\eqref{System 1} is positive, it follows from Proposition~\ref{Proposition 3} that~\eqref{Theorem 1-0-1} holds. Moreover, according to Lemma~\ref{Lemma 2}, there exists vector $\v >\bnull$ such that~$\v>\x^{\star}$ and that~\eqref{Lemma 2-2} holds. Let $\x(t,\x_0)$ be the solution of the undelayed system~\eqref{Theorem 1-0-0} corresponding to the initial condition $\x(0)=\x_0$. The proof will broken up into three steps:

\begin{enumerate}
\item[$\mathbf{i)}$] First, we show that the solutions $\x(t,\bnull)$ and $\x(t,\v)$ of~\eqref{Theorem 1-0-0} starting from the initial conditions $\x(0)=\bnull$ and $\x(0)=\v$, respectively, converge to $\x^{\star}$ as $t\rightarrow \infty$.
\item[$\mathbf{ii)}$] Second, we prove that for any non-negative initial condition $\x_0\in\mathbb{R}^n_+$, there exists a vector $\overline{\v}>\bnull$ such that $\overline{\v}\geq \x_0$ and~\eqref{Lemma 2-2} holds for $\overline{\v}$.
\item[$\mathbf{iii)}$] Finally, we show that for any $\x_0\in\mathbb{R}^n_+$, the solution $\x(t,\x_0)$ of~\eqref{Theorem 1-0-0} converges to $\x^{\star}$ as $t\rightarrow \infty$.
\end{enumerate}

\textbf{Step $\mathbf{i)}$} Since $\f+\g$ is cooperative, it follows from~\cite[Proposition 3.2.1]{Smith:95} that $\x(t,\bnull)$ is non-decreasing and $\x(t,\v)$ is non-increasing for all $t\geq 0$, which implies that
\begin{align*}
\bnull \leq \x(t,\bnull)\leq \x(t,\v)\leq \v,\quad t\geq 0.
\end{align*}
Thus, $\x(t,\bnull)$ and $\x(t,\v)$ are bounded and monotone. It now follows from~\cite[Theorem 1.2.1]{Smith:95} that $\x(t,\bnull)$ and $\x(t,\v)$ converge to an equilibrium of~\eqref{Theorem 1-0-0} in $[0,\v]$, which means that $\x(t,\bnull),\x(t,\v)\rightarrow \bar{\x}^{\star}$, where
\begin{align*}
\f(\bar{\x}^{\star})+\g(\bar{\x}^{\star})=\bnull.
\end{align*}
We claim that $\bar{\x}^{\star}=\x^{\star}$. By contradiction, suppose this is not true. Then, it is easy to verify that
\begin{align*}
\x(t)=\bar{\x}^{\star}\neq \x^{\star},\quad \forall t\in [-\tau_{\max},\infty),
\end{align*}
satisfies~\eqref{System 1}. This shows that for the non-negative initial condition $\bphi_{\bar{x}^{\star}}(t)=\bar{\x}^{\star}$, $t\in[-\tau_{\max},0]$, the solution $\x(t,\bphi_{\bar{x}^{\star}})$ of~\eqref{System 1} does not converge to $\x^{\star}$ $(\x(t,\bphi_{\bar{x}^{\star}})=\bar{\x}^{\star}\neq \x^{\star}$ for all $t\geq 0)$, contradicting the fact that $\x^{\star}$ is the globally asymptotically stable equilibrium of~\eqref{System 1}. Therefore, $\x(t,\bnull)$ and $\x(t,\v)$ converge to $\x^{\star}$ as $t\rightarrow \infty$.

\textbf{Step $\mathbf{ii)}$} Let $\x_0\in\mathbb{R}^n_+$ be an arbitrary initial condition and let $\v>\bnull$ be a vector satisfying~\eqref{Lemma 2-2}. Then, we can choose $\gamma\geq 1$ such that $\x_0\leq \gamma\v$. Define $\overline{\v}=\gamma\v$. As $\f$ and $\g$ are sub-homogeneous, we have
\begin{align*}
\f(\overline{\v})+\g(\overline{\v})&\leq \gamma^\alpha\bigl(\f(\v)+\g(\v)\bigr)< \bnull,
\end{align*}
where the right-most inequality follows from~\eqref{Lemma 2-2}.

\textbf{Step $\mathbf{iii)}$} According to the previous step, for any initial condition $\x_0\in\mathbb{R}^n_+$, we can find a vector $\overline{\v}>\bnull$ such that $\overline{\v}\geq \x_0$ and that~\eqref{Lemma 2-2} holds for $\overline{\v}$. As $\f+\g$ is cooperative, system~\eqref{Theorem 1-0-0}  is monotone~\cite[p. 34]{Smith:95}, which implies that
\begin{align*}
\x(t,\bnull)\leq \x(t,\x_0)\leq \x(t,\overline{\v}), \quad \forall t\geq 0.
\end{align*}
From Step $\mathbf{i)}$,  we have $\x(t,\bnull),\x(t,\overline{\v})\rightarrow \x^{\star}$  as $t\rightarrow \infty$, and hence $\x(t,\x_0)$ also converges to $\x^{\star}$. Hence,~\eqref{Theorem 1-0-0} is globally asymptotically stable for all non-negative initial conditions.

%
%

\subsection{Proof of Lemma~\ref{Lemma 2}}\label{Lemma:2}

We first show that if system~\eqref{System 1} has a globally asymptotically stable equilibrium at $\x^{\star}\in\mathbb{R}^n_+$, then $(a)$ holds.

$(a)$ Assume that there is vector $\w\neq \x^{\star}$ such that $\w\geq \x^{\star}$ and that~\eqref{Lemma 2-1} holds. Define $\bphi_w(t)=\w$ and $\bpsi_w(t)=\w$, $t\in[-\tau_{\max},0]$. According to Lemma~\ref{Lemma 3}, we have
\begin{align}
\y(t,\bpsi_w)\leq \x(t,\bphi_w), \quad \forall t\geq 0,
\label{Lemma 2-5}
\end{align}
where $\x(t,\bphi_w)$ and $\y(t,\bpsi_w)$ are solutions of~\eqref{System 1} and~\eqref{System 1-1}, respectively. Moreover,  according to~\cite[Corollary 5.2.2]{Smith:95}, $\y(t,\bpsi_w)$ is non-decreasing for all $t\geq 0$, implying that
\begin{align}
\w\leq\y(t,\bpsi_w), \quad \forall t\geq 0.
\label{Lemma 2-6}
\end{align}
It follows from~\eqref{Lemma 2-5} and~\eqref{Lemma 2-6} that $\w\leq \x(t,\bphi_w)$ for all $t\geq 0$. Therefore,
$\x(t,\bphi_w)\nrightarrow \x^{\star}$, contradicting the fact that $\x^{\star}$ is globally asymptotically stable.

$(b)$ According to part $(a)$, we have
\begin{align*}
(\f+\g)(\w)\ngeq \bnull,\quad \forall\w\geq\x^{\star},\;\w\neq \x^{\star}.
\end{align*}
Since $\f+\g$ is cooperative, it follows from~\cite[Proposition 4.2]{Vahid:11} that there is some vector~$\v>\x^{\star}$ satisfying~\eqref{Lemma 2-2}.

%
%

\bibliographystyle{IEEEtran}
\bibliography{bibliografia}

\end{document}